\documentclass[aps,prb,superscriptaddress,floatfix,twocolumn,longbibliography]{revtex4-2}

\usepackage{amsmath,dcolumn,bm,amsthm,tabularx,subfigure}
\usepackage[colorlinks=true,urlcolor=blue,citecolor=blue,linkcolor=blue]{hyperref}
\usepackage{mathrsfs,times}
\usepackage{bbold,dsfont}
\usepackage{graphicx,graphics,color}
\usepackage{mathtools,nicefrac}
\usepackage{slashed}
\usepackage{amsfonts,amssymb}
\usepackage{multirow}
\usepackage{booktabs}
\allowdisplaybreaks

\newcommand{\beq}{\begin{equation}}
\newcommand{\eeq}{\end{equation}}
\newcommand{\bea}{\begin{eqnarray}}
\newcommand{\eea}{\end{eqnarray}}

\newcommand{\R}{{\mathcal{R}}}
\newcommand{\Lft}{{\mathcal{L}}}

\begin{document}

\title{Inner structure of the many-body localization transition and the fulfillment of the Harris criterion}

\author{Jie~Chen}
\thanks{These authors contributed equally to this work}
\email[\\Contact author:\ ]{chenjie666@xhu.edu.cn}
\affiliation{Key Laboratory of Artificial Structures and Quantum Control (Ministry of Education), School of Physics and Astronomy, Shanghai Jiao Tong University, Shanghai 200240, China}
\affiliation{School of Science, Key Laboratory of High Performance Scientific Computation, Xihua University, Chengdu 610039, China}

\author{Chun~Chen}
\thanks{These authors contributed equally to this work}
\email[\\Contact author:\ ]{chunchen@sjtu.edu.cn}
\affiliation{Key Laboratory of Artificial Structures and Quantum Control (Ministry of Education), School of Physics and Astronomy, Shanghai Jiao Tong University, Shanghai 200240, China}

\author{Xiaoqun~Wang}
\email[Contact author:\ ]{xiaoqunwang@zju.edu.cn}
\affiliation{School of Physics, Zhejiang University, Hangzhou 310058, Zhejiang, China}
\affiliation{Collaborative Innovation Center of Advanced Microstructures, Nanjing University, Nanjing 210093, China}

\date{\today}

\begin{abstract}

We treat the disordered Heisenberg model in 1D as the standard model of many-body localization (MBL). Two new and independent order parameters stemming solely from the half-chain von Neumann entanglement entropy $S_{\textrm{vN}}$ are introduced to probe the eigenstate phase transition in this model. From the symmetry-endowed entropy decomposition, they are the probability distribution deviation $|d(p_n)|$ and the von Neumann entropy $S_{\textrm{vN}}^{n}(D_n\!=\!\mbox{max})$ of the maximally dimensional symmetry subdivision. The finite-size scaling reveals that $\{p_n\}$ drives the localization transition, preceded by a thermalization breakdown transition governed by $\{S_{\textrm{vN}}^{n}\}$. For the noninteracting case, these transitions coincide, but in the interacting circumstance they separate. Such separability creates an intermediate phase regime and discriminates between the Anderson and MBL transitions. One obstacle whose solution eludes the community to date concerns the violation of the Harris criterion in most numerical investigations of MBL. Upon elucidating the mutually independent measures comprising $S_{\textrm{vN}}$, it becomes clear that the previous studies may lack the resolution to pinpoint thus potentially overlook the crucial internal structure of the transition. We show that after this necessary decomposition, the universal critical exponents for both transitions of $|d(p_n)|$ and $S_{\textrm{vN}}^{n}(D_n\!=\!\mbox{max})$ fulfill the Harris criterion: $\nu\approx2\ (\nu\approx1.5)$ for quench (quasirandom) disorder. Our work puts forth symmetry combined with entanglement as an organization principle for the generic eigenstate matter and phase transition.

\end{abstract}

\maketitle

%\tableofcontents

\section{Introduction}

In the noninteracting systems, quenched disorder causes the Anderson localization of the single-particle wavefunctions \cite{anderson1958absence,Basko2,gornyi2005interacting}. While, in the presence of weak interaction, MBL is expected to occur in the many-body wavefunctions \cite{oganesyan2007localization,pal2010many}. The wavefunction's entanglement naturally plays an important role in both these nonequilibrium phenomena. Within the traditional thermalization, an isolated quantum system eventually reaches the uniform energy and particle distribution, but in MBL, certain parts of the system remain localized, not participating in the energy propagation or state mixing \cite{nandkishore2015many,abanin2019colloquium}. Preservation of the long-time coherence of quantum bits employing MBL can be a vital step in constructing the scalable quantum computers and quantum memories.

The Heisenberg spin chain with quenched disorder is often referred to as the standard model in the MBL research, and has been extensively studied for decades \cite{vznidarivc2008many,luitz2015many}. However, presently there remain some significant controversies in several key aspects of the problem \cite{vsuntajs2020quantum,de2016absence,DeRoeck,lev2015absence,vznidarivc2016diffusive,agarwal2017rare,sierant2020polynomially,morningstar2022avalanches,kiefer2020evidence,khemani2017two,zhang2018universal}. Prominently, a large portion of the numerical evaluations of the MBL transition therein (assumed continuous) yield the critical exponents near 1 \cite{luitz2015many,khemani2017two,kjall2014many}, violating the Harris criterion \cite{harris1974effect}. Although there exist a few interesting exceptions \cite{NicoKatz,Gray,ZHSun} detailed below at the end of Sec.~\ref{sepmbl}, this discrepancy is widely recongized as a bottleneck in the exact diagonalization (ED) approach to MBL, whose mechanism remains elusive for quite some time. Commonly, researchers would attribute it to the system-length limitation or the finite-size effect \cite{khemani2017two,zhang2018universal}. Here, we provide a different perspective to interpret this discrepancy.

The early seminal works on MBL \cite{Basko2,oganesyan2007localization} have won the Onsager prize in 2022. But ever since the appearance of \cite{vsuntajs2020quantum} in 2020, the status of MBL becomes somehow undermined. Community exhibits the growing skepticism about its existence. Focus exclusively on the eigenvalue properties, \cite{vsuntajs2020quantum} advocated the global long-range spectral correlations or structures (based on the Thouless-Heisenberg time ratio) over the local short-range spectral measures (based on the level-spacing ratio) in assessing the MBL transition from the ergodic-phase side in the disordered Heisenberg chain. Surprisingly, \cite{vsuntajs2020quantum} pointed out that this standard model of MBL might not accommodate MBL because the critical disorder strength thereof drifts to infinity linearly as the chain length increases. Interestingly, as one recent reaction to \cite{vsuntajs2020quantum}, \cite{morningstar2022avalanches} proposed a more comprehensive theoretical framework, where at least four different looking landmarks were imposed across the so-called MBL regimes. Critically, \cite{morningstar2022avalanches} targeted the avalanche instability \cite{DeRoeck} of the MBL transition in the disordered Heisenberg chain via coupling it to an infinite heat bath where the feedback was prohibited. Upon explicitly breaking the system's internal U(1) symmetry, the companion open-system Lindbladian simulation in \cite{morningstar2022avalanches} produced a continuously varying thermalization rate proposed to constrain the lower bound of the transition in the closed systems. Moreover, regarding the many-body resonances, a natural physical picture is that the systemwide resonances (corresponding to the nonlocal transports due to the quantum superpositions) should be first suppressed by weak disorder. The eventual entrance to the MBL phase at strong disorder is then controlled by the local deterministic (or classical) transport bottlenecks in finally achieving the ceasing of the nearest-neighbor exchanges or communications across the bonds. But this reasoning seems not to immediately consist with \cite{morningstar2022avalanches}.  

\begin{figure}[tb]	
	\begin{center}		
		\includegraphics[width=1\linewidth]{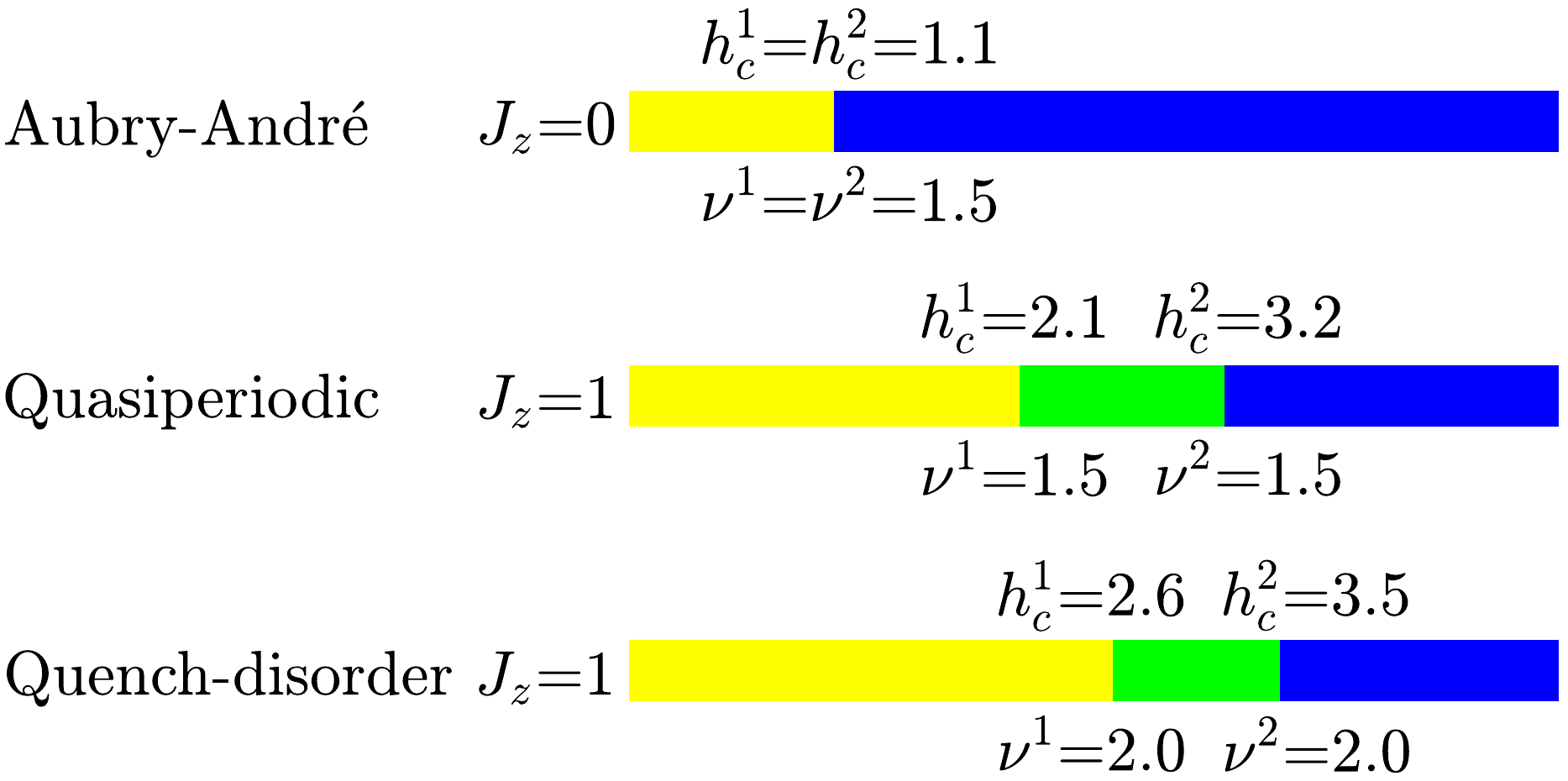}		
		\caption{In the three prototype scenarios, the Heisenberg chain (\ref{ham}) randomized in the diagonal field strength displays two disentangled transition points $h^{1,2}_c$, each characterized by their own critical exponents $\nu^{1,2}$.}
		\label{phase_all}			
	\end{center}
\end{figure}

In view of the intensified controversy, the goal of the present work is to alert that the paradigmatic MBL transition in the disordered Heisenberg chain might possess the peculiar inner structures enriched by the U(1) symmetry. The organization of these hidden structures is most transparently seen from the entanglement of the many-body wavefunctions. Historically, the Harvard experiment \cite{lukin2019probing} was among the first to demonstrate that when the quantum system possessed the U(1) symmetry, it was possible to decompose its entanglement entropy into the pieces of the number and the configurational entropies. In this work, exploiting the very concept of the entanglement entropy decomposition and extending it to the study of the eigenstate phase transition, two new and mutually independent order parameters are devised. As shown by Fig.~\ref{phase_all}, they each govern a phase transition point. Besides uncovering an in-between phase regime, this decomposition and the companion finite-size analyses yield respectively two critical exponents both satisfying the Harris criterion. To our knowledge, such a fundamental inner structure of the MBL transition has not been reported before.

The paper is structured as follows. In Sec.~\ref{model} we introduce the model and explain in detail the procedure of the symmetry-resolved entanglement decomposition. In Sec.~\ref{twomeasures} we define the two entanglement measures deployed throughout this work and stress the importance of their mutual independence. As a benchmark test, in Sec.~\ref{aamodel} we examine the appropriateness of the proposed methodology in the noninteracting quasirandom model and find the consistency of the approach with the known results. Section~\ref{sepmbl} is the central part of the present work where we implement the entropy decomposition strategy to study the MBL transition zone in the representative randomized interacting models. Particularly, we find the two-transition-point structure and the separation between the thermalization breakdown transition and the localization transition controlled respectively by the two independent entanglement measures we introduced. Section~\ref{experiment} proposes how to detect the two entanglement measures in the real experiments and demonstrates the relevance of a measurable correlator in aiding the location of the thermalization breakdown transition. Section~\ref{sepmbl} gives the conclusion. To reach larger system sizes, in Appendix~\ref{app1} we adopt the shift-invert method to recheck our results presented in Sec.~\ref{sepmbl} up to $L=20$ and find good agreement. Appendix~\ref{app2} presents the additional results for the scaling of the symmetry-resolved entanglement entropy in the blocks with the submaximal dimension. It turns out the corresponding results are consistent with that obtained from the block with the maximal dimension. Finally, Appendix~\ref{app3} shows the robustness of the Harris criterion and the proposed entropy decomposition tactic when they are applied to analyze the MBL transitions in the randomized interacting model with the Gaussian-distributed type disorder. 

\section{Model and entropy decomposition} \label{model}

The open spin-$\frac{1}{2}$ $X\!X\!Z$ chain is described by the following Hamiltonian,
\begin{equation}
H=\sum^{L-1}_{i=1}{[ J( S_{i}^{x}S_{i+1}^{x}+S_{i}^{y}S_{i+1}^{y}) +J_{z}S_{i}^{z}S_{i+1}^{z}]}+\sum^{L}_{i=1}{h_i}S_{i}^{z}.
\label{ham}
\end{equation}
Upon the Jordan-Wigner transformation, model (\ref{ham}) is equivalent to the spinless-fermion Hubbard model in 1D,
\begin{align}
H=&\sum^{L-1}_{i=1}{[-t( c_{i}^{\dagger}c_{i+1}+\textrm{H.c.})+\frac{U}{2}(2n_{i}-1)(2n_{i+1}-1)]} \nonumber \\
&+\sum^{L}_{i=1}{h_i}(2n_{i}-1),
\label{hamfermion}
\end{align}
with the substitutions $-t\!=\!2J$ and $\frac{U}{2}\!=\!J_z$. Using the hard-core bosons, model (\ref{ham}) is simultaneously equivalent to the following Bose-Hubbard-like model,
\begin{align}
H=&\sum^{L-1}_{i=1}{[\frac{J}{2}( b_{i}^{\dagger}b_{i+1}+\textrm{H.c.})+J_z(\frac{1}{2}-b^\dagger_ib_i)(\frac{1}{2}-b^\dagger_{i+1}b_{i+1})]} \nonumber \\
&+\sum^{L}_{i=1}{h_i}(\frac{1}{2}-b^\dagger_ib_i).
\label{hamhardcore}
\end{align}
To some extent, this multifacetedness reflects the conceptual importance of $H$. In the quasiperiodic case, we set $h_i\!=\!h\cos(2\pi\xi i+\phi)$ where $\xi\!=\!\sqrt{2}$ and $\phi\!\in\![-\pi,\pi)$. While, in the quench-disorder case, we choose $h_i\!\in\![-h,h]$ randomly from a uniform distribution. In this work, we mainly set $J\!=\!J_z\!=\!1$ to work with the isotropic Heisenberg model. When $J_z\!=\!0$, the model is noninteracting and if the modulation is quasiperiodic, (\ref{hamhardcore}) is called the Aubry-Andr\'{e} model. Given the U(1) symmetry, we can fix the total magnetization $\sum^{L}_{i=1}{S_{i}^{z}\!=\!0}$ and impose the open boundary conditions throughout the calculations. Physically, this amounts to dealing with the 1D disordered Fermi-Hubbard model at the half-filling.

Below, let us explain the entropy decomposition \cite{lukin2019probing,goldstein2018symmetry,xavier2018equipartition,bonsignori2019symmetry,parez2021quasiparticle,kiefer2020evidence,luitz2020absence,kiefer2021slow,feldman2019dynamics,jiechen2021,jiechen2023} in more detail. We first show that this decomposition is mathematically exact. Then, we provide the physical picture for this decomposition to explain why it comprises a framework to study the key physical properties of the system. It is worth noting again that this entropy decomposition was first proposed and successfully implemented in the Harvard experiment \cite{lukin2019probing}, suggesting that the associated concepts and quantities of this procedure were not artificial but measurable in real laboratories. 

When the system respects the U(1) symmetry, meaning that the system's Hamiltonian commutes with the system's total number operator, $[H,N]=0$, then when the system is divided into the equal-length left and right parts, the half-chain reduced density matrix of a simultaneous eigenstate of operators $H$ and $N$ of the system will be block diagonal, $\rho=\oplus_n\rho_{n}$. This crucial relation can be proved as follows. Assume that the many-body wavefunction $|\psi\rangle$ is an eigenstate of $N$ and $H$, then, by definition, it is ready to show that
\begin{equation}
[N,|\psi\rangle\langle\psi|]=[N_{\mathcal L},|\psi\rangle\langle\psi|]+[N_{\mathcal R},|\psi\rangle\langle\psi|]=0,
\label{comm1}
\end{equation}
where $N_{\mathcal L}$ and $N_{\mathcal R}$ denote the total number operators of the left and right half-chains. Next, by tracing out the degrees of freedom on the right half-chain on both sides of Eq.~(\ref{comm1}), one obtains that
\begin{align}
{\rm Tr}_{\mathcal R}\{[N_{\mathcal L},|\psi\rangle\langle\psi|]\}+{\rm Tr}_{\mathcal R}\{[N_{\mathcal R},|\psi\rangle\langle\psi|]\}=0.
\label{comm2}
\end{align}
As $N_{\mathcal L}$ only acts on the left half-chain, one finds that
\begin{align}
{\rm Tr}_{\mathcal R}\{[N_{\mathcal L},|\psi\rangle\langle\psi|]\}=[N_{\mathcal L},{\rm Tr}_{\mathcal R}\{|\psi\rangle\langle\psi|\}].
\label{comm3}
\end{align}
Likewise, by definition,
\begin{align}
{\rm Tr}_{\mathcal R}\{[N_{\mathcal R},|\psi\rangle\langle\psi|]\}&=\sum_{i_{\mathcal R}}\langle i_{\mathcal R}|\bm(N_{\mathcal R}|\psi\rangle\langle\psi|-|\psi\rangle\langle\psi|N_{\mathcal R}\bm)|i_{\mathcal R}\rangle \nonumber \\
&=0,
\label{comm4}
\end{align}
where $\{|i_{\mathcal R}\rangle\}$ forms a complete set of the orthonormal basis states for the right half-chain. Plug Eqs.~(\ref{comm3}) and (\ref{comm4}) into Eq.~(\ref{comm2}), one obtains the exact relation that
\begin{align}
[N_{\mathcal L},{\rm Tr}_{\mathcal R}\{|\psi\rangle\langle\psi|\}]=[N_{\mathcal L},\rho_{\mathcal L}]=0.
\label{comm5}
\end{align}
As usual, the reduced density matrix on the left half-chain is defined by
\begin{equation}
\rho_{\mathcal L}={\rm Tr}_{\mathcal R}\{|\psi\rangle\langle\psi|\}.
\end{equation}
Therefore, Eq.~(\ref{comm5}) guarantees that the above reduced density matrix has a block diagonal structure,
\begin{align}
\rho_{\mathcal L}=\oplus^N_{n=0}\rho_{{\mathcal L},n},
\end{align}
where $n=0,1,2,...,N$ represents the possible eigenvalue of operator $N_{\mathcal L}$, and $\rho_{{\mathcal L},n}$ represents the block reduced density matrix within $\rho_{\mathcal L}$ that is built upon those basis states where the left half-chain accommodates exactly $n$ particles. Clearly, similar results can also be obtained for $\rho_{\mathcal R}$.

To properly introduce the entanglement entropy for each block density matrix, one needs to factor out the normalization coefficient
\beq
p_n={\rm Tr}_{{\mathcal L},n}\rho_{{\mathcal L},n}.
\label{eqnpn}
\eeq
Because $\sum_n{\rm Tr}_{{\mathcal L},n}\rho_{{\mathcal L},n}={\rm Tr}_{{\mathcal L}}\rho_{{\mathcal L}}=1$, $p_n$ is the probability of getting $n$ in the projective measurement of $N_{\mathcal L}$ upon $|\psi\rangle$. Now, one can introduce the normalized block reduced density matrix, $\tilde{\rho}_{{\mathcal L},n}=\frac{\rho_{{\mathcal L},n}}{p_n}$, satisfying ${\rm Tr}_{{\mathcal L},n}\tilde{\rho}_{{\mathcal L},n}=1$, and the total reduced density matrix becomes $\rho_{\mathcal L}=\oplus^N_{n=0}p_n\tilde{\rho}_{{\mathcal L},n}$. The total von Neumann entanglement entropy can now be rewritten as follows,
\begin{align}
S_{\rm vN}&=-{\rm Tr}_{\mathcal L}\{\rho_{{\mathcal L}}\log_2\rho_{{\mathcal L}}\} \nonumber \\
&=-\sum^N_{n=0}{\rm Tr}_{{\mathcal L},n}\{[p_n\tilde{\rho}_{{\mathcal L},n}]\log_2[p_n\tilde{\rho}_{{\mathcal L},n}]\} \nonumber \\
&=-\sum^N_{n=0}\left(p_n\log_2 p_n\right)-\sum^N_{n=0}\left(p_n{\rm Tr}_{{\mathcal L},n}\{\tilde{\rho}_{{\mathcal L},n}\log_2\tilde{\rho}_{{\mathcal L},n}\}\right) \nonumber \\
&=S_N+S_C.
\label{eqnsvn}
\end{align}
The above expression means that the total von Neumann entanglement entropy can be decomposed into the particle number entropy 
\begin{equation}
S_N=-\sum^N_{n=0}p_n\log_2p_n
\label{eqnsn}
\end{equation}
and the configurational entropy
\begin{equation}
S_C=\sum^N_{n=0}p_nS^n_{\textrm{vN}},
\label{eqnsc}
\end{equation}
where the block von Neumann entropy, also termed the symmetry-resolved entanglement entropy, assumes
\begin{equation}
S^n_{\textrm{vN}}=-{\rm Tr}_{{\mathcal L},n}\{\tilde{\rho}_{{\mathcal L},n}\log_2\tilde{\rho}_{{\mathcal L},n}\}.
\label{eqnsvnn}
\end{equation}
As can be seen, this entropy decomposition depends only on the U(1) symmetry and the definition of the entanglement entropy, thus it is mathematically exact, valid for both noninteracting and interacting, finite and infinite systems.

Next, we describe the physical picture for this decomposition. On the one hand, $S^n_{\rm vN}$ is a partial configuration entropy that measures the entanglement between the left and right half-chains, originating from the state or configuration superpositions due to particles' hoppings and interactions without changing the index $n$. Thus, $S^n_{\rm vN}$ quantifies the nonlocal fluctuations and correlations encoded in $|\psi\rangle$ away from the entanglement cut (the middle bond of the chain). Further, each $S^n_{\rm vN}$ carries a probability weight $p_n$; the sum of their product forms the total configuration entropy $S_C$. On the other hand, $S_N$ stands for the entanglement between the left and right half-chains, arising from the particle number fluctuations across the entanglement cut with the filling faction of the system fixed. Thus, $S_N$ quantifies the local fluctuations and correlations in $|\psi\rangle$ right at the entanglement cut. The above entropy decomposition naturally distinguish between these two different contributions. We think that it uncovers the necessary finer structures that help clarify the essential physical mechanisms behind the eigenstate transitions in generic U(1)-symmetric quantum systems. Moreover, as already mentioned, the recent Harvard experiment \cite{lukin2019probing} has successfully measured both $S_N$ and the configuration correlators that are somehow equivalent to $S_C$ or $S^n_{\rm vN}$ (see Fig.~\ref{main_pic4}), and found the area law of $S_N$ and the volume law of the configuration correlators in time evolving a state within the MBL phase or regime. In sum, this entropy decomposition is essential to the study of the MBL transition. We think that it is not an artifact of the method itself.

\section{Independence between the two measures} \label{twomeasures}

In Ref.~\cite{jiechen2023} and also from the explicit derivations in Eqs.~(\ref{eqnsvn})-(\ref{eqnsc}) above, we first point out that $p_n$ and $S_{\textrm{vN}}^{n}$ defined in Eqs.~(\ref{eqnpn}) and (\ref{eqnsvnn}) represent the genuine entanglement-based measures for the generic eigenstate matter, thanks to the importance of their mutual independence. The mutual independence of these two entanglement measures hints at the potential existence of the two transition points. In this regard, the use of the total $S_{\textrm{vN}}$ or the mutually dependent set $\{S_N,S_C\}$ can be problematic. By resort to the eigenstate properties of $p_n$ and $S_{\textrm{vN}}^{n}$, our intent here is to explore the internal structures of the ergodicity breaking transition in the Heisenberg-type models enabled by the improved resolution from symmetry combined with entanglement. To this end, we first elaborate on the proper definitions for the pair of the independent measures built purely upon the quantum entanglement.

For the absolute probabilities $p_n$, we calculate the inter-block probability deviation by measuring how far the actual block probabilities deviate from their ideal thermal distribution. This deviation is expressed as:
\begin{equation}
\left| d(p_n) \right|=\Big\{\!\sum_n \left| p_n - p_{n}^{\textrm{ideal}} \right|\!\Big\}_\textrm{av}.
\label{dpn}
\end{equation}
Here, \lq\lq av'' indicates the averaging over the samples and the eigenstates. For all the quantities involved, we use $5000$ random samples to perform the averages at $L=8,10,12,14$, and use $3000$ random samples for the averages at $L=16$. The ideal thermalization means that all the states are ergodic and have the equal accessibility probabilities. So the ideal block probabilities $p_{n}^{\textrm{ideal}}=\frac{D_n}{D}$ are determined by the ratios of the block dimensions relative to the total Hilbert-space dimension. For systems of length $L$ and the total particle number $N$, the relevant dimension $D_{N}^{L}=\frac{L!}{N!(L-N)!}$. Further, with a half-filled occupancy, $D=D_{L/2}^{L}$ and $D_n=D_{n}^{L/2}D_{L/2-n}^{L/2}$, so $p_{n}^{\textrm{ideal}}=\frac{D_{n}^{L/2}D_{L/2-n}^{L/2}}{D_{L/2}^{L}}$.

\begin{figure*}[t]
\begin{center}
\includegraphics[width=1\linewidth]{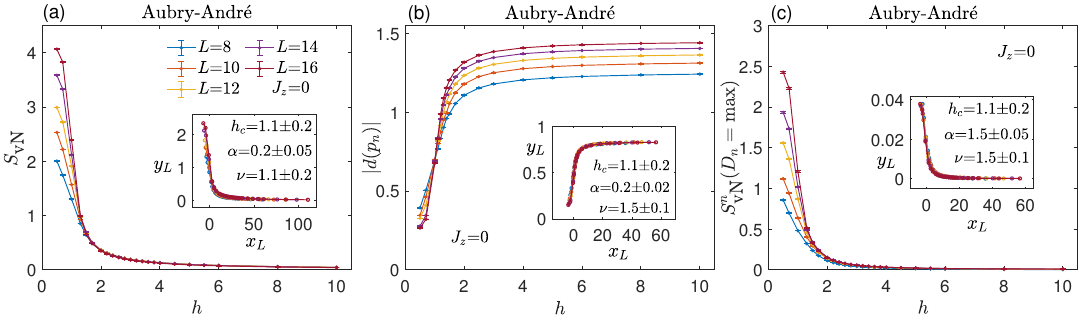}
\caption{The finite-size scaling of $S_{\textrm{vN}}$, $|d(p_n)|$, and $S^n_{\textrm{vN}}(D_n=\max)$ for the noninteracting Aubry-Andr\'{e} model $(J_z=0)$.}
\label{main_pic2}
\end{center}
\end{figure*}

For the relative intra-block von Neumann entanglement entropy $S_{\textrm{vN}}^{n}$, it is evident that in a finite chain, not all blocks are equivalent; the most significant block suffering less from the finite-size limitations corresponds to the one with the greatest dimension. In light of the fact that the block dimensions are largest in the middle and smaller on the sides, we adapt the convention that when $N=\frac{L}{2}$ is even, the central block with $n=\frac{N}{2}$ has the largest dimension; while, when $N=\frac{L}{2}$ is odd, the dimensions of the central blocks $\frac{N+1}{2}$ and $\frac{N-1}{2}$ are the same, so their contributions are averaged. Accordingly, we define the second independent measure as the entropy of the block with the maximal dimension:
\begin{equation}
S_\textrm{vN}^{n}(D_n\!=\!\max)\!=\!\left\{\begin{array}{c}
\{S_\textrm{vN}^{N/2}\}_{\textrm{av}}\ \ \ \ \ \ \ \ \ \ \ \ \ \ \ \ \ \ \ \ \ \ \ \ \ \ \ \ \ N\text{ even}\\[0.1em]
\!\!\{\frac{1}{2}[S_\textrm{vN}^{\frac{1}{2}(N+1)}+S_\textrm{vN}^{\frac{1}{2}(N-1)}]\}_{\textrm{av}}\ \ \ N\text{ odd}\\
\end{array}\!.\right.
\label{svnn}
\end{equation}
Afterward, we will average the pertinent quantities over all the available samples and eigenstates with the explicit indication omitted for brevity.

\section{Benchmark on the Aubry-Andr\'{e} model} \label{aamodel}

According to the general finite-size scaling theory, the scaling ansatz for the target quantity may take the following functional form \cite{kjall2014many}:
\begin{equation}
Q(L,h) =g(L) f\bm{(}(h -h _c) L^{1/\nu}\bm{)}.
\label{scaling}
\end{equation}
Here, we set $g(L)=L^\alpha$ and introduce the rescaled variables as $y_L = Q(L,h)/L^\alpha$ and $x_L = (h-h_c)L^{1/\nu}$. The critical strength $h_c$ and the critical exponent $\nu$ can then be extracted by collapsing the raw data onto a smooth curve of $x_L$ vs $y_L$.

As a benchmark test, Fig.~\ref{main_pic2} presents the numerical results for the noninteracting quasiperiodic Aubry-Andr\'{e} (AA) model. It is well known that due to the self-dual symmetry, 1D AA model features a modulation-strength tuned eigenstate phase transition between the extended and the localized states in an eigenenergy independent way, whose critical point is at the self-dual point $\frac{h}{J}=1$ \cite{aubry1980analyticity,Roosz}. Executing the above-introduced new strategy, Figs.~\ref{main_pic2}(b),(c) first depict the respective evolution of $|d(p_n)|$ and $S_{\textrm{vN}}^{n}(D_n\!=\!\mbox{max})$ as a function of $h$ for a series of chain lengths. The companion data collapse in the insets delivers an estimate of $h_c\approx1.1$ (close to the exact result) for both the critical points, suggesting that the two transitions overlap in the AA model and this inseparability might be protected by the self-dual symmetry. Following instead the conventional approach, Fig.~\ref{main_pic2}(a) shows the evolution of the total entanglement entropy $S_{\textrm{vN}}$ for the same localization transition where the same critical point is obtained as expected. Strictly, one shall always use $|d(p_n)|$ and $S_{\textrm{vN}}^{n}(D_n\!=\!\mbox{max})$ to probe the eigenstate properties of the AA model, as they yield the more accurate critical exponents $\nu\approx1.5$ for the better satisfaction of the Harris-Luck bound \cite{harris1974effect,luck1993classification}.

Based on the scheme of entropy decomposition, we provide the physical interpretations to explain why these two transitions coincide at the same parameter point in noninteracting models. For the noninteracting circumstances, we see from Eq.~(\ref{hamfermion}) that due to $U=0$, both $|d(p_n)|$ and $S_\textrm{vN}^{n}(D_n\!=\!\max)$ are generated from the single-particle hopping term. The difference is that $|d(p_n)|$ is controlled by the single-particle hopping connecting the left half-chain with the right half-chain via the entanglement cut (i.e., the middle bond of the chain). But, $S_\textrm{vN}^{n}(D_n\!=\!\max)$ is controlled by those single-particle hoppings within both the left and the right half-chains. In the absence of the interactions, these individual single-particle tunnelings are mutually independent and uncorrelated. Therefore, adding the quasiperiodic randomness changes simultaneously both these two entanglement measures at the same pace. This explains why the two transitions would typically occur at the same critical disorder strength for the noninteracting quasiperiodic systems. This reasoning consists with the scaling results of the noninteracting AA model presented in Fig.~\ref{main_pic2}.

\begin{figure*}[t]
	\begin{center}
		\includegraphics[width=1\linewidth]{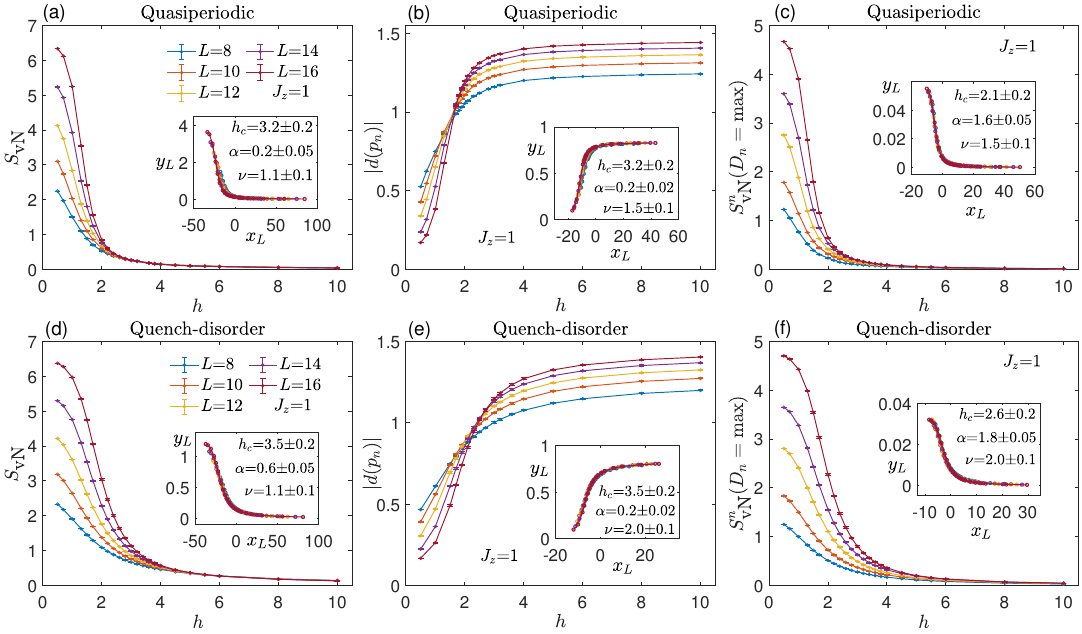}
		\caption{Upper (lower) row: The finite-size scaling of $S_{\textrm{vN}}$, $|d(p_n)|$, and $S^n_{\textrm{vN}}(D_n=\max)$ for the quasiperiodically (quench) disordered Heisenberg spin chain model, where the interaction strength $J_z=1$ is uniform and finite.}
		\label{main_pic3}
	\end{center}
\end{figure*}

To circumvent the Harris criterion, recently there is a surge of interest to invoke the BKT scaling ansatz to collapse the ED data of the MBL transition \cite{vsuntajs2020quantum}, which is inspired by the RG calculations \cite{Goremykina,Dumitrescu}. Three comments are in order. First, there seems to be a logic gap between the two approaches: Motivated by \cite{Goremykina,Dumitrescu}, \cite{vsuntajs2020quantum} favored the BKT ansatz and relied on it to suggest the absence of MBL. But, in the RG analyses of \cite{Goremykina,Dumitrescu}, the proposal of the BKT scaling law itself was built upon the assumption that MBL exists and is stable in the first place. Secondly, it is known that the numerical probe of the BKT scaling law is notoriously hard even for the phenomenological RG calculations \cite{Goremykina,Dumitrescu}. We tend to suspect that the ED based finite-size numerics could lead to any reliable conclusions regarding the BKT ansatz. Thirdly, the use of (\ref{scaling}) in the AA model is well accepted \cite{Roosz}, and from the analogous continuous trends perceived from Figs.~\ref{main_pic2} and \ref{main_pic3}, we attempt to continue our scaling analyses of the obtained ED results based still on (\ref{scaling}). 

\section{Separability of the MBL transition} \label{sepmbl}

It is worth emphasizing that the existence of the two phase transitions roots in the U(1) symmetry and the independence of the entanglement measures, thus being mathematically exact, valid for both noninteracting and interacting circumstances, and suffering no finite-size limitations in practice. Particularly, the utility of this proposed decomposing tactic manifests when being applied to assess the highly debated ergodicity breaking transition in the many-body systems. Using the finite-size numerics, below we first show that unlike the AA model, the interaction effects generically drive the separation between the two eigenstate transition points. More encouragingly, we further demonstrate that both their respective critical exponents derived from the same finite-size scaling analyses appear to meet the Harris criterion.

For the quasiperiodic modulation, we set $J_z=1$ and repeat the ED calculation across the same parameter range of $h$. Figures~\ref{main_pic3}(a)-(c) illustrate that as compared to the AA model (see Fig.~\ref{main_pic2}), the resulting critical points of $|d(p_n)|$ and $S_{\textrm{vN}}^{n}(D_n\!=\!\mbox{max})$ shift respectively to $3.2$ and $2.1$, thus being widely separated. It is a common feature that the transition of $|d(p_n)|$ typically occurs after the transition of $S_{\textrm{vN}}^{n}(D_n\!=\!\mbox{max})$, indicating that the entrance to the MBL phase is set by the local particle transport rather than the nonlocal configuration superposition. Moreover, the critical point drawn from $S_\textrm{vN}$ is also largely identical to that of $|d(p_n)|$. By contrast, the various critical exponents drawn from the data collapse appear to be insensitive to the rise of the interaction strength.     

In the case of random fields in the presence of many-body interaction $(J_z=1)$, as evidenced by Figs.~\ref{main_pic3}(d)-(f), we also observe the unambiguous separation of the two transition points akin to the interacting quasiperiodic case. Concretely, we find that the localization transition of $|d(p_n)|$ happens at $h_c\approx3.5$, in line with the established estimate in the literature based on $S_{\textrm{vN}}$. Furthermore, at $h_c\approx2.6$, we identify the distinct transition of $S_{\textrm{vN}}^{n}(D_n\!=\!\mbox{max})$ associated to the disruption of the thermalization. Remarkably, when computing the critical exponents for these two distinguishable transitions, we obtain the same value $\nu\approx2$ for both transition points, which is twice bigger than the exponent 1.1 drawn from $S_{\textrm{vN}}$, thereby satisfying the Harris-Chayes-Chayes-Fisher-Spencer (CCFS) bound \cite{harris1974effect,chayes1986finite,chayes1989correlation}.

Now we try to explain why the separation of these two transitions is a generic feature of randomized, interacting systems with U(1) symmetry. When the interactions are included into the model, these previously independent single-particle processes become correlated, building up some kind of rigidity that is immune to stronger disorder strengths. For our purpose, these interaction-induced correlations might still be divided into two categories: the local correlations concentrated around the entanglement cut and the nonlocal correlations extended more into the left and right half-chains. Physically, the scaling of $|d(p_n)|$ is controlled by the local correlations between the two half-chains near the entanglement cut, while the scaling of $S_\textrm{vN}^{n}(D_n\!=\!\max)$ is controlled by the nonlocal correlations between the two half-chains away from the entanglement cut. Generally, one would expect that the nonlocal correlations, being delicate, are more fragile or sensitive to the disorder than the local correlations. Therefore, under the increase of disorder strength, the transition point of $S_\textrm{vN}^{n}(D_n\!=\!\max)$ usually proceeds the transition point of $|d(p_n)|$. In other words, the randomness first suppresses the nonlocal transport arising from the coherent quantum-state superposition; after that, the local deterministic (or classical) transport bottlenecks are then achieved. Namely, the complexity and fluctuations, signature of thermalization, are first reduced within each individual block of the reduced density matrix, and then the particle transport across the local bonds involving different blocks is ceased to enter the MBL regime. The above elaboration provides the potential physical picture to interpret why the two transitions of $|d(p_n)|$ and $S_\textrm{vN}^{n}(D_n\!=\!\max)$ in the disordered, interacting systems are generically separated. Moreover, combine the above reasoning with the results shown in Fig.~\ref{main_pic3}, we tend to think that this qualitative feature of the transition-point separation is independent of the disorder types. It happens in both types of quench-disorder (boxed uniform and Gaussian distributed) as well as quasiperiodic, interacting models.

The physical nature of this intermediate phase regime in between the separated thermalization breakdown and localization transitions is a crucial problem in its own right. A thorough investigation of this topic will possess the potential to help us better understand the underlying mechanism and the subtle structures of the putative MBL transition or crossover. In view of the fact that this study is beyond the scope of current work, we tend to leave it to a follow-up research project headed toward this important direction \cite{ChenIntermediate}. Essentially, armed with the general symmetry decomposition principle, we will look deeper into the organization and evolution patterns of the entanglement spectrum drawn from the reduced density matrix for the low-dimensional disordered many-body systems. Moreover, besides the static diagnostics, we also pay attention to the dynamics of quantum information and treat it as another indispensable tool to clarify the physical nature of this intermediate phase regime. Incidentally, from Fig.~\ref{main_pic3}, we notice that this intermediate phase regime persists for both the quasiperiodic and quench disorder models. Phenomenologically, they look similar. As there is no rare-region Griffiths effect, thus no avalanche instability, in the quasiperiodic many-body systems. It is uncertain about the role of the avalanche instability in this intermediate phase regime as well as in the transition zone for the case of the quench disorder.

Below we conduct a search in the literature and select several representative papers that also studied the MBL transition employing the probes such as the level statistics and entanglement entropy including both their means and variances, the indicators from eigenstate correlations and bipartite fluctuations, as well as the eigenstate's Schmidt gap, the decohered entanglement entropy, and the diagonal entropy. Further, we also comment on the discrepancy between the critical exponents obtained respectively from the numerical simulations and the phenomenological RG treatments of the MBL transition. Particularly, we highlight that the reason why past approaches failed might be owing to the unawareness or the inability to distinguish these multi transition-point structures, resulting in the potential inaccuracy of the obtained smaller critical exponents. Moreover, we stress that the definitions of our proposed measures $|d(p_n)|$ and $S^n_{\rm vN}(D_n={\rm max})$ are unique and mathematically sound, thus they help resolve the problems of some prior works arising from the complete ignorance of the off-diagonal elements or the great amount of the eigenvalues of the reduced density matrix. We hope that this clarification of the relation between our work and the previous literature could strengthen the impact of the symmetry-resolved entropy decomposition scheme on the MBL community.

Those previous studies conceived various quantities to target the same transition zone but fell short of this bound. In the influential work \cite{luitz2015many}, a wide range of estimates based on level statistics and entanglement entropy including both their means and variances \cite{khemani2017two,kjall2014many}, as well as indicators from eigenstate correlations and bipartite fluctuations were jointly utilized to probe the same GOE-Poisson transition in model (\ref{ham}). Upon a thorough finite-size scaling analysis, Ref.~\cite{luitz2015many} suggested a consistent transition exponent $\nu\approx0.83$ across the entire dynamical phase diagram of (\ref{ham}). Ever since, more and more ED studies of MBL transitions encounter the likewise violation of the Harris-CCFS criterion. This might be owing to the unawareness or the inability to distinguish these multiple transition points, resulting in the potential inaccuracy of the obtained smaller critical exponents.

Facing this critically open question, besides the aforementioned proposal of using the BKT scaling law to avoid the need for satisfying the Harris criterion, there are few ED works claimed to devise the alternative quantities whose scaling behaviors tend to meet the Harris-CCFS bound. Especially, Ref.~\cite{Gray} advocated to employ the eigenstate's Schmidt gap defined by the modulus of the difference between the first two largest eigenvalues of the half-chain reduced density matrix to perform the finite-size scaling analysis across the transition zone. Interestingly, the critical exponent drawn from this procedure assumes $\nu\approx2.13$, thus rising above the bound. Likewise, Refs.~\cite{NicoKatz} and \cite{ZHSun} proposed to discard all the off-diagonal entries but use only the diagonal elements of the time-evolved density matrix of a quenched initial state to assess the MBL transition in model (\ref{ham}). Intriguingly, both the thus-yielded steady-state decohered entanglement entropy as introduced in \cite{NicoKatz} or the diagonal entropy as adapted in \cite{ZHSun} displayed the improved scaling trends that were consistent with the Harris-CCFS criterion.

Several caveats might need some attention here. Firstly, it was pointed out in \cite{Gray} that unlike the level statistics and entanglement entropy, the disorder-averaged Schmidt gap was not an order-parameter-like quantity but rather a scaling function for the MBL transition. Secondly, it is worth noticing that the diagonal entropy and the decohered entanglement entropy were both computational-basis-dependent \cite{ZHSun}, rendering their interpretations more specified. Thirdly, although being more feasible in experiments, the complete ignorance of the off-diagonal elements or the remaining eigenvalues of the density matrix lacks a clear physical justification and appears less controlled. In this sense, our proposed $\left| d(p_n) \right|$ and $S_\textrm{vN}^{n}(D_n\!=\!\max)$ in Eqs.~(\ref{dpn}) and (\ref{svnn}) could avoid the above-mentioned problems as their definitions are unique and mathematically sound.  

Incidentally, on the one hand, this violation of the Harris-CCFS criterion commonly occurs in the numerical ED studies of the MBL transition or crossover. But, on the other hand, most analytical results based on the phenomenological RG treatments of the MBL criticality show the good fulfillment of the same Harris-CCFS bound \cite{Vosk,Potter,zhang2018universal}. The system sizes used in the phenomenological RG are orders of magnitude longer than that in the ED studies at the cost that the microscopic details of the model are largely missing, let alone the symmetry-resolved entanglement decomposition. The origin of this discrepancy and how to reach a compromise between these leading approaches of MBL remain an outstanding question that needs further investigations.         

\begin{figure}[t]
\begin{center}
\includegraphics[width=0.75\linewidth]{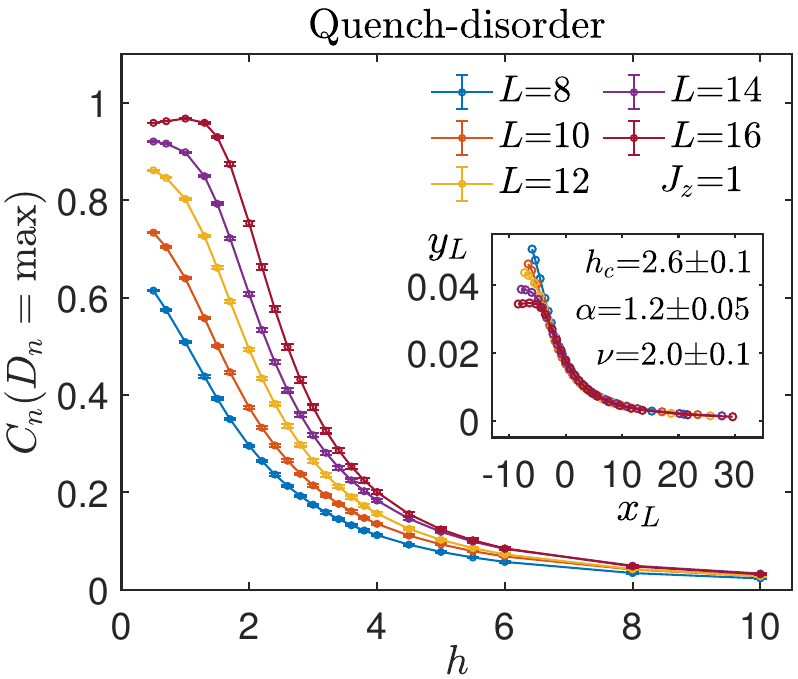}
\caption{The finite-size scaling of the correlator $C_n(D_n=\textrm{max})$ [see definition (\ref{cnmax})] as the experimentally more accessible alternative to detect the transition of $S^n_{\textrm{vN}}(D_n=\textrm{max})$ presented in Fig.~\ref{main_pic3}(f).}
\label{main_pic4}
\end{center}
\end{figure}

\begin{figure*}[t]
	\begin{center}
		\includegraphics[width=1\linewidth]{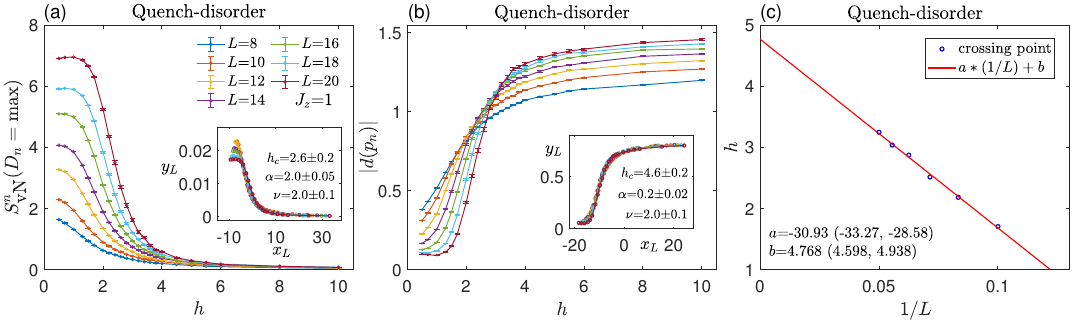}
		\caption{(a),(b) The finite-size scaling of $S_\textrm{vN}^{n}(D_n\!=\!\max)$ and $|d(p_n)|$ for the quench disordered Heisenberg spin chain model, where the interaction strength $J_z=1$ is uniform and finite. Here we use the shift-invert diagonalization method to reach the maximal system size $L=20$, and we focus exclusively on the eigenstate that is closest to the middle of the eigenspectrum. (c) displays the collected crossing points of the neighboring data lines in (b) as a function of the inverse system size $1/L$. The red solid line in (c) is a best linear fit for these available crossing points whose intercept $b$ represents an estimate for the critical disorder strength at the transition point in (b) in the infinite-$L$ limit.}
		\label{main_pic5}
	\end{center}
\end{figure*}

\section{Experimental feasibility} \label{experiment}

Both $p_n$ itself and a correlator
\begin{align}
C_n\!\coloneqq\!\sum_{\{\Lft_n\}}\!\sum_{\{\R_{N-n}\}}\!|p(\Lft_n\otimes \R_{N-n})-p(\Lft_n)p(\R_{N-n})|
\label{cnmax}
\end{align}
complementary to $S^n_{\textrm{vN}}$ were measured in the experiment \cite{lukin2019probing}. Here, $\{\Lft_n\}~(\{\R_{N-n}\})$ denotes all the possible configurations with $n~(N-n)$ particles in the left (right) half-chain. \cite{lukin2019probing} suggested that $C_n$ captures the qualitative characteristics of $S^n_{\textrm{vN}}$ through the quantification of the separability between $\Lft$ and $\R$. Indeed, by evaluating (\ref{cnmax}) for the disordered Heisenberg chain, we show in Fig.~\ref{main_pic4} that both the critical point and the critical exponent of the transition of $S^n_{\textrm{vN}}(D_n=\textrm{max})$ can be quantitatively reproduced via the scaling analysis of $C_n(D_n=\textrm{max})$. [Compare Figs.~\ref{main_pic3}(f) and \ref{main_pic4}.] In this regard, our predictions so far may largely be observable. Admittedly, here we assert the assumption that the eigenstate-averaged measurements we consider can be qualitatively implemented through the elaborate combinations of the disorder realizations and the initial state preparations for using the dynamics, as was successfully executed in the recent programmable-superconducting-processor-based experiment on MBL \cite{70}.

\section{Conclusion} \label{conclusion}

In this work, we introduce two new order parameters to characterize the MBL phase transition regime: the probability distribution deviation $|d(p_n)|$ and the entropy of the symmetry subdivision with the highest dimension $S_{\textrm{vN}}^{n}(D_n=\textrm{max})$, using the concept of entanglement entropy decomposition. Through finite-size scaling analyses, we find that $\{p_n\}$ drives the MBL phase transition, preceded by a thermalization breakdown phase transition governed by $\{S_{\textrm{vN}}^{n}\}$. For interacting systems, these transitions separate, but for noninteracting systems, they overlap. We obtain the universal critical exponents for these independent transitions: $\nu\approx1.5$ for the quasiperiodic-field case and $\nu\approx2$ for the random-field case. Notably, the latter exponent drawn from ED satisfies the Harris criterion. Finally, we demonstrate the experimental accessibility of these two probing order parameters via linking them to the probabilities, the correlation functions, and their variants that are measurable in laboratories.

\section{Acknowledgements}

This work is supported by the Ministry of Science and Technology of China (Grant No. 2016YFA0300500) and the NSF of China (11974244). C.~C. acknowledges the support from Shanghai Jiao Tong University startup fund and the sponsorship from Yangyang Development fund.

\appendix

\section{Longer-chain results from the shift-invert method} \label{app1}

To partly address the issue of the length limitation in ED, the shift-invert Lanczos method has been employed. This technique allows us to increase the maximal chain length solved in this work from $L=16$ up to $L=20$. Here we essentially revisit the classic work of Luitz et al. \cite{luitz2015many}. We use the same model and the same method, but the key improvement is that we implement the proposed entropy decomposition tactic to target the MBL transition zone. Incidentally, the tensor network methods for tackling the many-body systems are most suitable for characterizing phases or regimes with the area-law entanglement, so there are proposals like DMRG-X \cite{KhemaniPollmann} to acquire the excitation states deep inside the MBL regime. But, as the MBL transition zone accommodates various states exhibiting the volume-law entanglement, the tensor network approaches are relatively less employed to directly retrieve the eigenstates at the middle of the spectrum in this transition regime. 

Figure~\ref{main_pic5} presents the finite-size scaling results based on the two decomposed entanglement measures for the disordered Heisenberg spin chain on larger system sizes. Here we use $1000$ random samples for $L=20$, $3000$ samples for $L=18,16$, and $5000$ samples for other smaller $L$'s. We focus on the eigenstate closest to the middle of the eigenspectrum, namely the normalized energy of the target eigenstate is $\sim0.5$. From Fig.~\ref{main_pic5}(a) one can see that the scaling trend and the companion data collapse of $S_\textrm{vN}^{n}(D_n\!=\!\max)$ covering $L=8$ to $L=20$ deliver a critical exponent $\nu\approx2$ that is the same as that obtained in Fig.~\ref{main_pic3}(f) of the main text. But in Fig.~\ref{main_pic5}(a) we concentrate only on the middle of the spectrum---the position that is most difficult to be localized, therefore in principle the extracted critical disorder strength from Fig.~\ref{main_pic5}(a) should be greater or equal to the value obtained in Fig.~\ref{main_pic3}(f) where the contributions from the whole spectrum are included and averaged. Now, we find that the two estimates of the critical disorder strength for $S_\textrm{vN}^{n}(D_n\!=\!\max)$ are the same in both Fig.~\ref{main_pic5}(a) and Fig.~\ref{main_pic3}(f), suggesting that the putative mobility edge for the transition of $S_\textrm{vN}^{n}(D_n\!=\!\max)$ might be largely energy independent or only weakly depending on energy.

Likewise, upon the same procedure, we find from Fig.~\ref{main_pic5}(b) that the scaling trend and the accompanying data collapse of $|d(p_n)|$ covering $L=8$ to $L=20$ also deliver a critical exponent $\nu\approx2$, fulfilling the Harris-CCFS criterion. This estimate of the critical exponent is the same as that obtained from Fig.~\ref{main_pic3}(e) of the main text. However, here we notice that the critical disorder-strength estimate from the scaling of $|d(p_n)|$ is $h_c\approx4.6$ in Fig.~\ref{main_pic5}(b), which is larger than the corresponding estimate $h_c\approx3.5$ in Fig.~\ref{main_pic3}(e). This discrepancy hints that the putative mobility edge for the transition of $|d(p_n)|$ might accordingly exhibit a sharper energy dependence.

As the transition of $|d(p_n)|$ quantifies the transition to the MBL phase or regime, we perform a more careful determination of its critical disorder strength by using the crossing point of the two adjacent data lines of sizes $L-2$ and $L$ in Fig.~\ref{main_pic5}(b) to mark the estimated value of $h_c$ at the size $L$. After collecting all these available crossing points, we plot them as a function of $1/L$ in Fig.~\ref{main_pic5}(c). As shown there, we find that these crossing points tentatively follow a straight line in Fig.~\ref{main_pic5}(c). A linear fit gives an intercept $b=4.768$ for the linear fitting line, which is equivalent to a good estimate of $h_c$ for the transition of $|d(p_n)|$ in the infinite-size limit, i.e., when $L=\infty$. Therefore, by the two complementary approaches, we obtain from the scaling of $|d(p_n)|$ the consistent estimate of the critical disorder strength, $h_c\approx4.6$-$4.8$, for marking the entrance to the MBL phase or regime. This value is greater than that reported in \cite{luitz2015many}.

\section{Finite-size scaling of the symmetry-resolved entanglement entropy from the submaximal blocks} \label{app2}

\begin{figure}[tb]
\begin{center}
\includegraphics[width=0.71\linewidth]{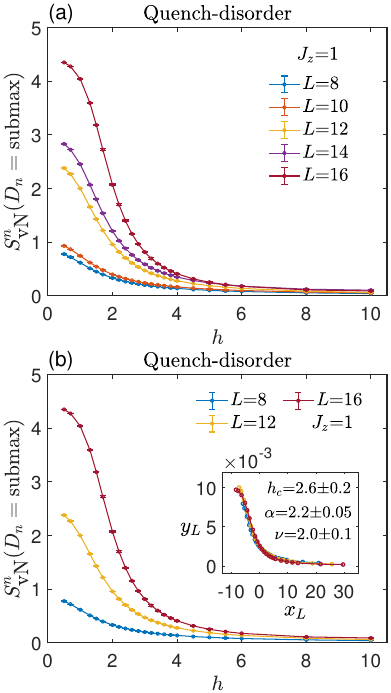}
\caption{(a) The finite-size scaling of $S^n_{\textrm{vN}}(D_n=\textrm{submax})$ for the quench disordered Heisenberg spin chain model, where the interaction strength $J_z=1$ is uniform and finite. (b) The data collapse of $S^n_{\textrm{vN}}(D_n=\textrm{submax})$ using only the selected data sets from $L=8,12,16$ to partly reduce the observed finite-size effect in (a).}
\label{main_pic6}
\end{center}
\end{figure}

We present the numerical results for the submaximal subspace next to the largest subspace and compare the findings with the results obtained for the largest subspace reported in the main text.

To this end, we introduce the symmetry-resolved entanglement entropy for the blocks with the submaximal dimension,
\begin{equation}
S_\textrm{vN}^{n}(D_n\!=\!\textrm{submax})\!=\!\left\{\begin{array}{c}
\{\frac{1}{2}[S_\textrm{vN}^{\frac{1}{2}(N+2)}+S_\textrm{vN}^{\frac{1}{2}(N-2)}]\}_{\textrm{av}}\ \ \ \ N\text{ even}\\[0.1em]
\!\!\{\frac{1}{2}[S_\textrm{vN}^{\frac{1}{2}(N+3)}+S_\textrm{vN}^{\frac{1}{2}(N-3)}]\}_{\textrm{av}}\ \ \ \ N\text{ odd}\\
\end{array}\!.\right.
\label{svnnsub}
\end{equation}
Figure~\ref{main_pic6}(a) shows the corresponding numerical results for the scaling of $S_\textrm{vN}^{n}(D_n\!=\!\textrm{submax})$ as that shown in Fig.~\ref{main_pic3}(f) of the main text for the scaling of $S_\textrm{vN}^{n}(D_n\!=\!\textrm{max})$. However, we notice that the finite-size effect of $S_\textrm{vN}^{n}(D_n\!=\!\textrm{submax})$ in Fig.~\ref{main_pic6}(a) is comparatively greater than that of $S_\textrm{vN}^{n}(D_n\!=\!\textrm{max})$ in Fig.~\ref{main_pic3}(f). Particularly, the lines of $L=8,10$ in Fig.~\ref{main_pic6}(a) are much closer to each other. The same trend happens also for the lines of $L=12,14$. Clearly, this enhanced finite-size effect prohibits us from performing a finite-size data collapse covering all the available data sets in Fig.~\ref{main_pic6}(a) from $L=8$ to $L=16$. As can be seen from Fig.~\ref{main_pic3}(f), this issue does not arise in the case of $S_\textrm{vN}^{n}(D_n\!=\!\textrm{max})$.

A qualitative way to understand the above observation can be as follows. One first assumes the clean case (i.e., $h=0$) and then use the dimension of the block, $\frac{N!}{n!(N-n)!}$, as a rough estimate of $S_\textrm{vN}^{n}(D_n\!=\!\textrm{submax})$. Next, one uses the ratio of the block dimensions between the two neighboring system sizes to quantify the closeness of the data lines. A simple calculation shows that the ratio exhibits an alternating pattern given by the two values of $\{2-\frac{3}{n+2},2+\frac{2}{n}\}$ with $n=2,3,4$ for $L=8$ to $L=16$. For instance, the ratio between $L=8,10$ equals $2-\frac{3}{2+2}=\frac{5}{4}$, while the ratio between $L=10,12$ equals $2+\frac{2}{2}=3$, suggesting a much larger gap between the lines of $L=10,12$ than that between the lines of $L=8,10$. This is consistent with the results shown in Fig.~\ref{main_pic6}(a). Therefore, we find that this finite-size effect will eventually be gone when the length $L$ (i.e., $n$) becomes sufficiently large. The issue arising here is due to the limitation of the small system lengths available in the ED approach.

By the same token, we find that for the case of $S_\textrm{vN}^{n}(D_n\!=\!\textrm{max})$, the block-dimension ratio changes to the following two values: $\{2-\frac{1}{n+1},2\}$. Now, the ratio between $L=8,10$ equals $2-\frac{1}{2+1}=\frac{5}{3}$, while the ratio between $L=10,12$ equals $2=\frac{6}{3}$. Therefore, in sharp contrast, even at small system lengths, the gap between the data lines of $S_\textrm{vN}^{n}(D_n\!=\!\textrm{max})$ for the neighboring $L$'s shows almost negligible finite-size effect. This allows us to perform the data collapse of $S_\textrm{vN}^{n}(D_n\!=\!\textrm{max})$ in Fig.~\ref{main_pic3}(f) using all the available chain lengths.

As an expedient way to partially address the finite-size issue in $S_\textrm{vN}^{n}(D_n\!=\!\textrm{submax})$, we perform a more restricted finite-size data collapse in Fig.~\ref{main_pic6}(b) using only the data sets of $S_\textrm{vN}^{n}(D_n\!=\!\textrm{submax})$ from $L=8,12,16$. It turns out that by this way, we get the same critical disorder strength $h_c\approx2.6$ and the same critical exponent $\nu\approx2.0$ from the data collapse of $S_\textrm{vN}^{n}(D_n\!=\!\textrm{submax})$ as that obtained from the data collapse of $S_\textrm{vN}^{n}(D_n\!=\!\textrm{max})$, hinting at the internal consistency of the type of entanglement measures we advocate.

\begin{figure}[tb]
	\begin{center}
		\includegraphics[width=0.71\linewidth]{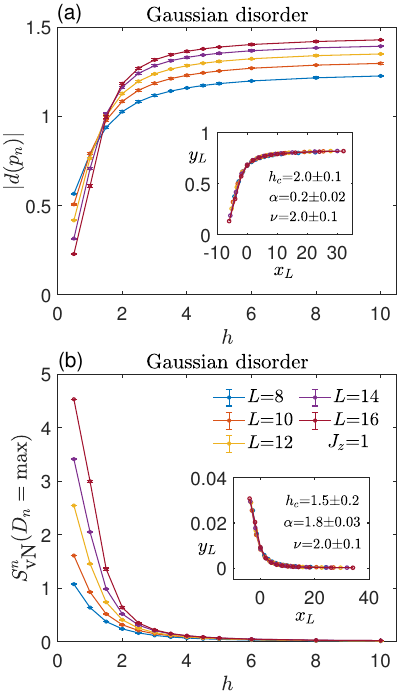}
		\caption{The finite-size scaling of $|d(p_n)|$ and $S_\textrm{vN}^{n}(D_n\!=\!\max)$ for the quench disordered Heisenberg spin chain model, where the interaction strength $J_z=1$ is uniform and finite, and the onsite disorder strength $h_i$ is now drawn from a Gaussian distribution as given by Eq.~(\ref{gauss}). The inset in each panel displays the corresponding finite-size data collapse, and the obtained critical parameter estimates are explicitly indicated.}
		\label{main_pic7}
	\end{center}
\end{figure}

\section{Model of the Gaussian distributed quench disorder} \label{app3}

To investigate the sensitivity and robustness of the Harris-CCFS criterion as well as the entanglement decomposition strategy to other disorder configurations, we change the disorder type from the box distribution to the Gaussian distribution and repeat the same numerical finite-size scaling analyses based on the independent entanglement measures $|d(p_n)|$ and $S_\textrm{vN}^{n}(D_n\!=\!\max)$ for the Gaussian randomized Heisenberg spin chain model.

It is known that in a uniform box distribution, the probability of generating a number between $x$ and $x+{\rm d}x$ equals
\begin{equation}
p(x){\rm d}x=\!\left\{\begin{array}{c}
{\rm d}x\ \ \ \ \ x\in[-h,h]\\[0.1em]
0\ \ \ \ \ \textrm{otherwise}\\
\end{array}\!.\right.
\end{equation}
When switching to the Gaussian distribution, the probability becomes
\beq
p(x){\rm d}x=\frac{1}{\sqrt{2\pi}h}e^{-x^2/(2h^2)}{\rm d}x,
\label{gauss}
\eeq
where we have chosen the mean to be zero and the standard deviation to be the tunable disorder strength, i.e., $\sigma=h$. One salient difference between the box and Gaussian distributions is that the range of the random number is bounded in the box distribution, but in the Gaussian distribution, the random number is unbounded: it is mostly concentrated around the mean but can be any real number with an exponentially suppressed probability controlled by the variance $\sigma^2$.

Figure~\ref{main_pic7} shows the respective evolutions of $|d(p_n)|$ and $S_\textrm{vN}^{n}(D_n\!=\!\max)$ as a function of the disorder strength $h$ for the modified disordered Heisenberg chain model (\ref{ham}) where the onsite randomized field strength $h_i$ is drawn from the Gaussian distribution as per Eq.~(\ref{gauss}). As can be seen there, both the critical exponents extracted from the finite-size data collapses of $|d(p_n)|$ and $S_\textrm{vN}^{n}(D_n\!=\!\max)$ equal $2$, being the same as that obtained for the case of the box distributed quench disorder [see Figs.~\ref{main_pic3}(e),(f)]. This result suggests that provided that the disorder is independent, uncorrelated or short-range correlated, and the variance of its distribution is not diverging, then for the putative continuous MBL transitions, the Harris-CCFS criterion should generically be the same, namely the critical exponent $\nu\geqslant2/d$, regardless of the disorder types or the detailed forms of the distributions. Moreover, we speculate that the proposed finite-size scaling analyses of the transition zone based upon the procedure of the entanglement decomposition work equally well for other disorder types or different nonuniform disorder distributions. In addition, we notice that due to the fact that some values of $h_i$ in the Gaussian distribution can be much larger than the standard deviation $h$, the critical disorder-strength values $h_c$'s from $|d(p_n)|$ and $S_\textrm{vN}^{n}(D_n\!=\!\max)$ in Fig.~\ref{main_pic7} are both smaller than their counterparts in the case of the box distribution [see Figs.~\ref{main_pic3}(e),(f)]. But, the separation of the two-transition structure remains observable in this type of the Gaussian distributed quench disorder.

\bibliography{ref}

\end{document}